\documentclass[12pt,preprint]{aastex}

\def\mearth{ M_{\earth}}
\def\rearth{ R_{\earth}}

\begin{document}

\title{Design Considerations for a Ground-based Transit Search for Habitable Planets
Orbiting M dwarfs}

\author{Philip Nutzman and David Charbonneau \altaffilmark{1}}
\affil{Harvard-Smithsonian Center for Astrophysics, 60 Garden St., 
Cambridge, MA 02138}

\altaffiltext{1}{Alfred P. Sloan Research Fellow}
\email{pnutzman@cfa.harvard.edu}
\keywords{stars: low mass --- stars: planetary systems --- techniques:
photometric --- surveys --- extrasolar planets}

\begin{abstract}

By targeting nearby M dwarfs, a transit search using modest equipment is 
capable of discovering planets as small as 2 $R_{\earth}$ in the habitable zones of their host stars.  
The MEarth Project, a future transit search, aims to employ a
network of ground-based robotic telescopes to monitor  
M dwarfs in the northern hemisphere with sufficient precision and cadence to detect such planets.
Here we investigate the design requirements for the MEarth Project.
We evaluate the optimal bandpass, and the necessary field of view, telescope aperture, and telescope time allocation 
on a star-by-star basis, as is possible for the well-characterized nearby M
dwarfs.  Through these considerations, 1,976 late M dwarfs ($R 
< 0.33 ~R_{\odot}$) emerge as favorable targets for transit monitoring.
Based on an observational cadence and on total telescope time allocation tailored
to recover 90 \% of transit signals from planets in habitable zone orbits, 
we find that a network of ten 30 cm telescopes 
could survey these 1,976 M dwarfs in less than 3 years.  A null
result from this survey would set an upper limit (at 99 \% confidence) of 17
\% for the rate of occurrence of planets larger than 2 $R_{\earth}$ in the
habitable zones
of late M dwarfs, and even stronger constraints for planets lying closer than the
habitable zone.  If the true occurrence rate
of habitable planets is 10 \%, the expected
yield would be 2.6 planets.

\end{abstract}

\section{Introduction}

In upcoming years, the study and characterization of exoplanets will
depend largely on the unique window that transiting planets offer into their
properties. Transit observations reveal a planet's radius, and in combination
with radial-velocity measurements, permit a determination of the planet's mass. This
combination of measurements provides the only available direct constraint on
the density and hence bulk composition of exoplanets.  When a planet cannot be
spatially resolved from its host star, transit-related observations typically
offer the only means for direct measurements of planetary emission and
absorption.
Already, transmission spectroscopy has probed the atmospheric chemistries of
HD~209458b
\citep{Charbonneau2002,Vidal-Madjar2003,Barman2007} and HD~189733b
\citep{Tinetti2007}, while infrared monitoring during secondary eclipse has led to
the detection of broadband thermal emission from HD 209458b, TrES-1, HD
189733b, HD 149026b, and GJ 436b \citep{Deming2005,Charbonneau2005,Deming2006,
Harrington2007,Deming2007}.
Precise spectroscopic measurements during secondary
eclipse have unveiled the infrared spectrum of HD 209458b
\citep{Richardson2007}, and HD 189733b \citep{Grillmair2007}.
Most recently, infrared observations gathered at a variety of orbital
phases have allowed the characterization of the longitudinal temperature
profiles for several Hot Jupiters 
\citep{Harrington2006,Knutson2007,Cowan2007}.
These and a host of other studies (as reviewed by Charbonneau et al 2007)
\nocite{Charbonneau2007} 
demonstrate the profound impact of the transiting class of exoplanets. 

Given the importance of the transiting planets, it is critical to extend
their numbers to planets in different mass regimes, different irradiation
environments, and around varied star types.  Until the recent
discovery that the Neptune-sized GJ 436b transits its host M dwarf \citep{Gillon2007}, 
all known transiting planets were hot gas giants orbiting
Sun-like stars.  A discovery like this points to the advantages
that M dwarfs provide for expanding the diversity of the transiting planets;
the advantages are particularly acute for the detection of rocky, habitable
planets (see e.g. Gould et al. 2003). \nocite{Gould2003}
We review these observational opportunities by explicitly considering the case of a 2 $R_{\earth}$ planet (representing
the upper end of the expected radius range for super-Earths, see Valencia et
al. 2007 and Seager et al. 2007) \nocite{Valencia2007,Seager2007} orbiting in
the habitable zones of the Sun and a fiducial M5 dwarf  ($0.25 ~M_{\odot},
0.25 ~R_{\odot}, 0.0055 ~L_{\odot}$). 
\begin{enumerate}
\item The habitable zones of M dwarfs are drawn in close to the
stars, improving the transit likelihood.  A planet receiving the
same stellar flux as the Earth would lie only 0.074 AU from the M5, and
would present a 1.6 \% geometric probability of transiting, 
compared to the 0.5 \% probability for the Earth-Sun system.  Note that here,
and throughout the paper, we define habitable zone orbits to be at the
orbital distance for which the planet receives the same insolation
flux as the Earth receives from the Sun. 
\item Transits from the habitable zones of M dwarfs happen much more
frequently.   At 0.074 AU from the M5, a planet would transit once every 14.5
days, compared to 1 year for the Earth-Sun system.  
This is critical for detectability, as dramatically less observational time is
required to achieve a transit detection.
\item The small radii of M dwarfs lead to much deeper transits.  The
2-$R_{\earth}$ planet would eclipse 0.5 \% of an M5's stellar disk area, but only
1 part in 3000 of that of the Sun. 
\item  The small masses of M dwarfs lead to larger induced radial
velocity variations. Taking a mass of $7 ~M_{\earth}$ for the super-Earth (in a
habitable zone orbit with P=14.5 days), the
induced peak-to-peak velocity variation on the M5 is 10 m/s, versus an induced 1.3 m/s
variation at 1 AU from the Sun. 
\end{enumerate}
A transit survey targeting nearby, proper motion selected M dwarfs would
also avoid a number of astrophysical false alarms
(see e.g. Mandushev et al. 2005 and O'Donovan et al. 2006).
\nocite{Mandushev2005,ODonovan2006}  Among the most common false alarms are
those caused by eclipsed giant stars, hierarchical triples composed of
a star plus an eclipsing pair, grazing eclipsing
binaries, and blends with fainter background eclipsing binaries.  
The survey would avoid eclipsed giant stars by construction; a giant
would never be confused as a nearby, high proper motion M dwarf.
Hierarchical triple systems would be exceedingly unlikely, 
given the red colors and low intrinsic luminosity 
of the system.  In any case, given its proximity, such a system would likely be
partially resolvable through high-resolution imaging.  
Grazing eclipsing binaries would also be unlikely, though interesting, given the rarity of double M
dwarf eclipsing pairs.  A spectroscopic study looking for
the presence of double lines could easily confirm or rule out this scenario.
Blends with background binaries are, in principle, still an issue for a targeted M dwarf
survey.  However, because of the high proper
motions of the M dwarfs, chance alignments could be confirmed or ruled out with archived 
or future high resolution observations.   

Aside from these observational advantages, several developments in 
astrophysics point to exciting possibilities with M dwarfs.
Firstly, the growing number of M dwarf exoplanet discoveries, including the
$\sim$5.5-$\mearth$ planet orbiting Gliese 581c \citep{Udry2007} and
the microlensing discovery OGLE 2005-BLG-390Lb \citep{Beaulieu2006}, 
suggest an abundance of \emph{sub}-Neptune mass planets orbiting M dwarfs.  It is an
open challenge to find a transiting planet in this mass regime; simply
obtaining a radius measurement for such a planet (for which there are no Solar
System analogs) would be extremely fruitful as it might allow one to distinguish between
rocky or ocean planet composition models \citep{Valencia2007}.  
Intriguingly, the idea that life can survive on habitable zone planets around M dwarfs has
been recently rehabilitated (see Scalo et al. 2007 and Tarter et al. 2007 for
detailed discussion) \nocite{Scalo2007,Tarter2007}.  Previously, it had been
assumed that the rotational synchronization expected of close-in habitable zone planets
would lead either to atmospheric collapse or to steep temperature gradients and
climatic conditions not suitable for life.  Works reviewed in Scalo et al. and Tarter et al.
argue that atmospheric heat circulation should prevent each of these
barriers to habitability.  Regardless, the absence of such heat
redistribution would be readily observable with precise infrared photometric
monitoring as a large day-night temperature
difference, while the detection of a small day-night difference
would provide a strong case for the existence of a thick atmosphere.  

With Spitzer and the James Webb Space Telescope (JWST), atmospheric
observations similar to those mentioned earlier for Hot Jupiters can be extended to habitable, Earth-sized planets orbiting M dwarfs.  This possibility is brought about by the small surface areas and
temperatures of M dwarfs, which lead to significantly more favorable 
planet-star contrast ratios.  This ratio for a habitable 2-$\rearth$ planet
orbiting an M5 is $0.05 ~\%$ (in the Rayleigh-Jeans limit), leading to secondary eclipse depths well reachable with Spitzer's sensitivity (this compared to a contrast ratio $0.0017 ~\%$ for a habitable 2-$\rearth$
orbiting the Sun).  JWST photometry will be capable of measuring the day-night
temperature difference for warm Earth-like planets orbiting M dwarfs
\citep{Charbonneau2007b}, thus addressing the extent of heat redistribution on
these planets and hence the presence or absence of an atmosphere.

It is interesting to consider the place of M dwarf planets in the expected
yields of ongoing and upcoming transit surveys.  The COROT \citep{Baglin2003}
and Kepler \citep{Borucki2003} space missions are the most ambitious of the transit
surveys; with long, uninterrupted time baselines and excellent
photometric precision, these missions should yield rocky planets with orbital
periods much longer than those detected by ground-based transit searches.
Gould et al. (2003) point out that missions like COROT and Kepler are much more  
sensitive to M dwarf habitable planets than to solar-type habitable planets, if
they can be reliably monitor the M dwarfs to faint magnitudes ($V > 17$).  In practice, 
stellar crowding, noise from sky background, and other technical issues
\citep{Gould2003,Deeg2004} strongly limit their sensitivity to these faint magnitudes.  
COROT and Kepler can precisely monitor bright, nearby M dwarfs, but
with Kepler observing one fixed field of roughly 100 square degrees, and
COROT monitoring much less sky area, these missions probe only a small
number of such nearby M dwarfs.   Wide-angle ground surveys, such as HATNet
\citep{Bakos2004}, SuperWASP \citep{Pollacco2006}, TrES
\citep{Alonso2004}, and XO \citep{McCullough2005}
cumulatively cover swaths of sky containing large numbers of nearby M dwarfs,
but at the expense of employing apertures too modest to effectively probe any but the brightest of
these M dwarfs (see e.g. McCullough \& Burke 2007).\nocite{McCullough2007}

Motivated by these difficulties, and the fact that the closest, most observationally favorable M dwarfs are spread sparsely throughout the sky, we consider an alternative approach in
which these M dwarfs are \emph{individually} targeted.   
In this paper, we develop a concept that we term the MEarth
Project, which envisions a cluster of robotic telescopes 
dedicated to targeted, sequential photometric monitoring of nearby M dwarfs.
We determine the necessary design elements for a survey searching for
transiting planets as small as $2~ R_{\earth}$ (the upper end of the rocky
planet regime), and out to the M dwarf habitable zones. 
In \S 2, we briefly discuss the MEarth Project concept. 
In \S 3 we describe already compiled lists of nearby M dwarfs suitable for
observations.  We discuss their observational properties and use these to
estimate basic stellar parameters.  In \S 4 we determine the necessary
telescope aperture through a calculation that estimates the photometric
precision for M dwarf stars.  In \S 5 we determine the necessary field of
view, which is driven by the need for a sufficient number of calibrator stars.  
In \S 6 we estimate the amount of gross telescope time necessary for a successful survey
for an optimal list of late M dwarf targets for the MEarth project.  
In \S 7 we wrap up with a discussion of our conclusions and the design
implications for the MEarth project.

\section{MEarth Project: Discussion}

The heart of the MEarth concept is to use a network of robotic telescopes 
to precisely monitor the brightness of roughly 2000 northern, nearby M dwarfs, with a
sensitivity sufficient to detect 2-$R_{\earth}$ planets.
The MEarth network will be housed in a single enclosure on Mt. Hopkins,
Arizona.  Multiple sites spread in longitude would
be observationally favorable, but would unfortunately be a cost-prohibitive arrangement.
The number of targets is selected to ensure that even a null result is
astrophysically interesting, while the sensitivity goal reaches into the upper
end of the radius range expected for rocky planets.  
If we take the fiducial M5V star 
as typical of the M dwarfs being monitored, and assume an
occurrence rate of 10 \% for habitable zone planets larger than 2 $\rearth$, the
expected yield from 2000 M dwarfs is 3.2 planets, which would complement Kepler's expected harvest of
habitable planets around Sun-like stars \citep{Gillon2005}.  
Correspondingly, a null result places an upper limit
for the occurrence rate of such habitable planets at 15 \% (at 99 \%
confidence).  Note that later we will refine this calculation using actual estimates of
$R_{\star}$ and $a_{\rm HZ}$ for 1,976 observationally favorable M dwarf targets. 

Perhaps the most critical aspect of the MEarth project is that the M dwarfs
are observed one-by-one.  This sequential mode of observing comes with a certain
benefit: the field of view requirements are relaxed and set only by
the need for the field to contain a sufficient number of comparison stars (see \S 5).
A modest field of view requirement opens up the possibility of using
off-the-shelf equipment and dodges many of the technical
challenges that beset wide-field transit searches (see e.g. Bakos et
al. 2004 and McCullough \& Burke 2007 for a discussion of these issues).

On the other hand, the observational cadence achieved per target when 
sequentially targeting M dwarfs is significantly less than when 
staring at and repeatedly imaging a single field.  There are two issues which
help compensate for the sparse cadence.  Firstly, typical levels and
timescales of correlated noise in photometric surveys \citep[see e.g.][]{Pont2006} 
suppress the benefit of dense time sampling such that the 
`standard' $N^{-1/2}$ improvement in precision generally does not apply. 
Secondly, the flexibility of being able to choose your targets and when to
observe them greatly enhances the efficiency per observation of the
transit survey.  We envision an adaptively scheduled transit search, 
wherein the observing sequence is updated as the images are
gathered and analyzed.  We consider a design in which transits
are identified while in progress by the automated reduction software.  The
subsequent alert triggers other telescopes in the MEarth array (or at another
observatory) for high-cadence monitoring at improved precision and in multiple
colors until a time after transit egress.
Intense coverage following this could then pin down the orbital period.  
Under this observing strategy, the amount of time required to achieve detection
is the amount of time until the first transit event falls during an observation
session.  This is significantly less time than is required for
current transit surveys, which spot transits in phase-folded archived data and
typically require at least 3 distinct transit events.

Note that in this adaptive mode of observing, a false positive triggered by
photometric noise is addressed immediately and, in most cases, easily
dismissed with a few
additional exposures.  False positives, in this context, are thus far less
costly than in traditional transit surveys.  This, coupled with the fact
that the MEarth network will monitor only a couple hundred M dwarfs on any given
night, means that the follow-up mode can be triggered at a relatively low
statistical threshold.  In our paper, we require a per-point photometric
precision that is three times smaller than the given transit depth of
interest.  A threshold near `3 sigma' would be outlandish for a traditional
transit survey monitoring hundreds of thousands of stars, but here would lead to
an inexpensive fraction of time spent on false alarms each night.  It is
important to note that while follow-up is triggered at relatively low
significance, a genuine transit would be detected by MEarth to higher
significance, as the entire MEarth network would be galvanized to high-cadence
follow-up.

The level of significance of this transit detection would depend on the
transit depth and on the details of the photometric noise, especially the
level of correlated noise on the timescale of a transit.  Most transit surveys
show such red noise at levels of 3-6 mmag for untreated light curves, which
can often be reduced to 1-2 mmag with decorrelation algorithms
\citep{Pont2007,Tamuz2005,Jenkins2000}.  As an example, the Monitor project
shows red noise levels of 1-1.5 mmag \citep{Irwin2007}. The
MEarth project's employment of multiple telescopes may be a weapon against red
noise if the systematics/correlated noise are largely independent from one
telescope to another.  Nevertheless, these considerations suggest that for
transit depths less than $\sim 5$ mmag, MEarth may have to alert an outside
observatory to achieve a very high significance transit detection.

\section{Catalog of Potential M dwarf Targets}

Despite their low intrinsic luminosities, M dwarfs are intrinsically abundant, 
and provide a bounty of bright survey targets.  We have consulted the
L\'epine-Shara Proper Motion Catalog of northern stars
\citep[LSPM-North;][]{LSPM2005} for 
potential targets and their observational properties.  
LSPM-North is a nearly complete list of northern stars with proper
motion greater than $0''.15 ~\mathrm{yr}^{-1}$. \citet{Lepine2005} identifies a
subsample of 2459 LSPM stars for which either trigonometric
parallaxes or spectroscopic/photometric distance moduli indicate that their
distance is less than 33 pc, as well as more than 1600 stars suspected to be dwarfs within 33 pc. 
Restricting to d $<$ 33 pc keeps the rate of contamination from high proper
motion subdwarfs small.  At this distance, incompleteness is mainly
due to the proper motion limit ($\mu > 0''.15 ~ \mathrm{yr}^{-1}$).

LSPM gives $V$ for stars with Tycho-2 magnitudes, and
estimates $V$ from USNO-B1.0 photographic magnitudes for the remaining stars. 
These magnitudes are supplemented with 2MASS $JHK$ magnitudes.
When there is a distance measurement in the literature, L\'epine
provides these (1676 with trigonometric parallaxes and 783 with
spectroscopic/photometric distance moduli).  For the remaining 1672 stars, L\'epine
estimates distance through a piecewise $V-J$ vs. $M_V$ relationship which is calibrated by stars with
known parallaxes.  In assigning distances, we always use the value tabulated
by L\'epine, except for a small fraction of cases when the trigonometric
parallax is uncertain by more than 15 \%.  In these cases we use L\'epine's
piecewise $V-J$ vs. $M_V$ relationship.
We cull the L\'epine (2005) subsample to probable M dwarfs
dwarfs by requiring $V-J> 2.3, J-K > 0.7, J-H > 0.15$ (motivated by
Figures 28 and 29 in L\'epine \& Shara 2005).  This leaves nearly 3300
probable nearby M, or late K, dwarfs. We hereafter refer to this culled
sample of stars as the LSPM M dwarfs.

\subsection{Estimating Stellar Parameters}

We considered three routes towards estimating the
luminosities, masses, and radii of the LSPM M dwarfs:
\begin{enumerate}
\item theoretical models of \citet{Baraffe1998}, which offer synthetic $V-K$
colors that can then be matched to observed
$V-K$ colors
\item empirically determined fits for the stellar parameters as a function
of $V-K$ colors 
\item the $M_{\star}$-$M_K$ relations of \citet{Delfosse2000},
combined with the empirical mass-radius relation of \citet{Bayless2006}, and the
bolometric corrections of \citet{Leggett2000}.  
\end{enumerate} 
The first method is problematic in that the theoretical models (not just those
of Baraffe et al. 1998) are known to underestimate radii by 5-15 \% for stars in the range 0.4
$M_{\odot} \lesssim $ M $ \lesssim 0.8 M_{\odot}$ \citep{Ribas2006}.
Furthermore, as noted by Baraffe et
al. (1998), the synthetic colors involving the $V$ band are systematically too blue
by $\sim 0.5 $ mags, which is suggested to be due to some unmodeled source of
$V$ band opacity.
While the second method avoids these problems,  
it suffers from significant dispersion in the stellar parameters for a
given $V-K$ (see for example Figure 2 of Delfosse et al. 2000). 
When the absolute $K$ magnitude, $M_K$, is well known, the third method does very
well at estimating the mass and radius, relying on the small intrinsic
scatter of the Delfosse and empirically determined
mass-radius relations. However, only a third of the LSPM M dwarfs have
trigonometric parallaxes, and the distance moduli for the remaining M dwarfs
have uncertainties up to $\pm$ 0.6 mag.  This uncertainty in distance
propagates to a roughly $ \pm 30 \%$ uncertainty in mass,
which is comparable to the scatter in the relations based on $V-K$ color. 
We settled on the third method, which at its worst produces errors
comparable to the second method, while performing significantly better when the
distance to the M dwarf is relatively well determined.

For each star, we insert the estimated $M_K$ into the polynomial fit of
Delfosse et al. (2000) to infer the mass.  We then apply a polynomial fit to
the mass-radius data of Ribas (2005) to convert this to a radius.  To estimate
the stellar luminosity, we adopt the bolometric corrections of Leggett (2000).
Given this luminosity and radius, we estimate the $T_{\mathrm{eff}}$, while we
combine the mass and radius to estimate the star's $\log{g}$. The determined
$T_{\mathrm{eff}}$ and $\log{g}$ drive our choice of synthetic spectra, as
described below in \S 4.1.1.

In Table 1, we show a selection of adopted stellar parameters for different radius bins,
along with approximate spectral types (calculated from the mean $V-K$ of each
bin, and using Table 6 of Leggett 1992).
We note that of all the estimated parameters, our calculations below 
are most sensitive to the inferred radius.  This is simply because 
the transit depth and hence the necessary photometric precision 
goes as $R_{\star}^{-2}$.  We estimate the uncertainty in radius for individual
determinations to be roughly 30-35 \% (though better than 15\% for the third
of stars with trigonometric parallaxes), with this figure dominated by the uncertainty in distance
modulus.  Errors at this level are tolerable (as long as they are not
significantly biased in one direction) and do not alter our conclusions.

\section{Telescope Aperture Requirements}
\subsection{Photometric Precision}

We follow standard calculations of photon, scintillation, and detector noise to
simulate photometric precision for a variety of possible observational set-ups.
The simulated systems described below are intended to be representative
of commercially available CCDs and telescopes.  We adopt the
site characteristics of the Whipple Observatory on Mt. Hopkins, Arizona 
(altitude 2350 m) and the specifications of common semi-professional
thinned, back-illuminated CCD detectors,
but allow for other parameters, such as the aperture and filter to vary.

We calculate the precision as follows:
\begin{equation}
\mathrm{precision} = \frac{\sqrt{N_{\star}+\sigma^2_{\mathrm{scint}}+n_{\mathrm{pix}} (N_S+N_D+N_R^2)}}
{N_{\star}}
\end{equation}
where $N_{\star}$ is the number of detected source photons, $n_{\mathrm{pix}}$ is the
number of pixels in the photometric aperture, $N_S$ is the number of photons
per pixel from background or sky, $N_D$ is the number of dark current
electrons per pixel, and $N_R$ is the RMS readout noise in electrons per
pixel. We adopt the scintillation expression of \citet{Dravins1998}
\begin{equation}
\frac{\sigma_{scint}}{N_{\star}} =  0.09 \frac{X^{3/2}}{D^{2/3} \sqrt{2 t}} \exp{(-
\frac{h}{8}})
\end{equation}
where $X$ gives the airmass (which we set at 1.5), $D$ gives the
aperture diameter in cm, $t$ gives the exposure time in s, and h
gives the observatory altitude above sea level in km.

For a common, commercially-packaged, Peltier-cooled CCD camera, $N_R$ =10
e$^-/$pixel, and $N_D$ = 0.1 e$^-/$pixel/s, each of which are negligible for bright
sources.  To calculate $n_{\mathrm{pix}}$ we assume a circular photometric
aperture of radius 5 $\arcsec$. For 13 \micron ~$\times$ 13 \micron ~pixels, and  
a typical focal ratio of f/8, $n_{\mathrm{pix}}= 62.9 * (D/30
~\mathrm{cm})^2$.   

We calibrated the sky background flux estimates (photons cm$^{-2}$ s$^{-1}$ arcsec$^{-2}$) 
for our simulated system using $i$ and $z$ band observations taken over several nights
with KeplerCam (see, e.g., Holman et al. 2007) and the 1.2 m telescope located at the Whipple Observatory.  
These measurements differ from what would be received by our hypothetical
system by a factor of the overall throughput
of the 1.2m system over the overall throughput of our system,
in each bandpass of interest.  We make a first order estimate of this ratio by
using KeplerCam observations of stars with calibrated i and z magnitudes to determine the
scale factor necessary for our simulated system to reproduce the number of
counts received by KeplerCam.  The sky background flux received by our
hypothetical system is then approximately the observed flux divided by
this scale factor. Of course,
the actual sky background present
in an exposure depends on many factors, in particular the phase of the moon
and its proximity to the target object.  For our calculations we take the
median of our sky flux measurements.  This estimate turns out to be an
overestimate of the true median sky value because a disproportionate fraction
of our observations were taken near full lunar phase. 

We calculate $N_{\star}$ with
\begin{equation}
N_{\star} = t \times \pi (D/2)^2 \times \int{T(\lambda) f(\lambda) \frac{\lambda}{h c} d\lambda}
\end{equation}
where t is the exposure time, D is the aperture diameter,
$f(\lambda)$ is the stellar flux described in \S 4.1.1 and $T(\lambda)$ is the
overall system transmission described in \S 4.1.2.

In our calculation, we do not include the potentially significant 
noise from the intrinsic variability of the star.  Though variability is
common among M dwarfs, it is often on timescales different from
the transit timescale or of a form distinct from a transit signal and thus removable.
One common form of M dwarf variability is that of flares, which are easily
distinguished from a transit in that flares result in an increase in flux as
opposed to a decrement.  Starspots are also a concern, but induce variability
on a timescale defined by the stellar rotation period, which is much longer
than that of a transit and hence may be distinguished. Because the MEarth
project is a targeted survey, troublesome variable M dwarfs could possibly be
dropped in favor of photometrically quiet M dwarfs (such as the
transiting-planet host GJ 436), though such variables might be worth retaining
for non-transit related studies.

\subsubsection{Synthetic M dwarf Spectra}

We employ \texttt{PHOENIX}/NextGen model spectra (see Hauschildt, Allard, \& Baron
1999 \nocite{Hauschildt1999}and references therein) to simulate the flux of our target M dwarfs.
These model spectra have subsequently been updated with new TiO and H$_2$O line
lists \citep{Allard2000}, which significantly improve M
dwarf spectral energy distributions on the blue side of the optical.  However, as
pointed out by \citet{Knigge2006}, this improvement appears to be somewhat at the
expense of accuracy in $I-K$ colors, which the original NextGen models reproduce well.  
Given the importance of this spectral
region for our studies, we exclusively use the original NextGen models.

The NextGen models are
available over the range of M dwarf temperatures (2000 K to 4000 K), in steps
of 100 K, for $ 4.0 \leq \log{g} \leq 5.5$ in steps of 0.5 dex, with [Fe/H]=0.  
For each star, we choose the spectrum with $T_{\mathrm{eff}}$ and $\log{g}$
most similar to the star's inferred $T_{\mathrm{eff}}$ and $\log{g}$.

The synthetic spectra give the star's surface flux, and therefore require a dilution
factor, $x=(R_{\star}/d)^2$, to reproduce the flux incident on the Earth's
atmosphere.  Since precise measurements of $R_{\star}$ and $d$ are not available for
the M dwarf candidates, we instead calculate $x$ by using observed
magnitudes and scaling with respect to the zero magnitude fluxes.  
For example, we calculate the $V=0$ dilution
factor by equating $f(\lambda)_{V}= \int T_{V}(\lambda) f(\lambda) d \lambda/\int
T_{V}(\lambda) d \lambda$ with the
zero point $f(\lambda)_{V=0}$ taken from Bessell and Brett (1988), where $T_V$ is the
standard bandpass response of $V$ (Bessell 1990, Bessell and Brett
1988).  

If the photometric colors of the star match well with the colors of the
synthetic spectrum, the dilution factor
depends little on which band is chosen for the calculation.  In practice, the
synthetic and observed colors do not necessarily match up well.
To more robustly estimate the dilution factor, we average the $x$'s determined
separately through the $V$,$J$, and $K$ bands.  

\subsubsection{Transmission}

\begin{figure}[t]
\epsscale{1.0}
\plotone{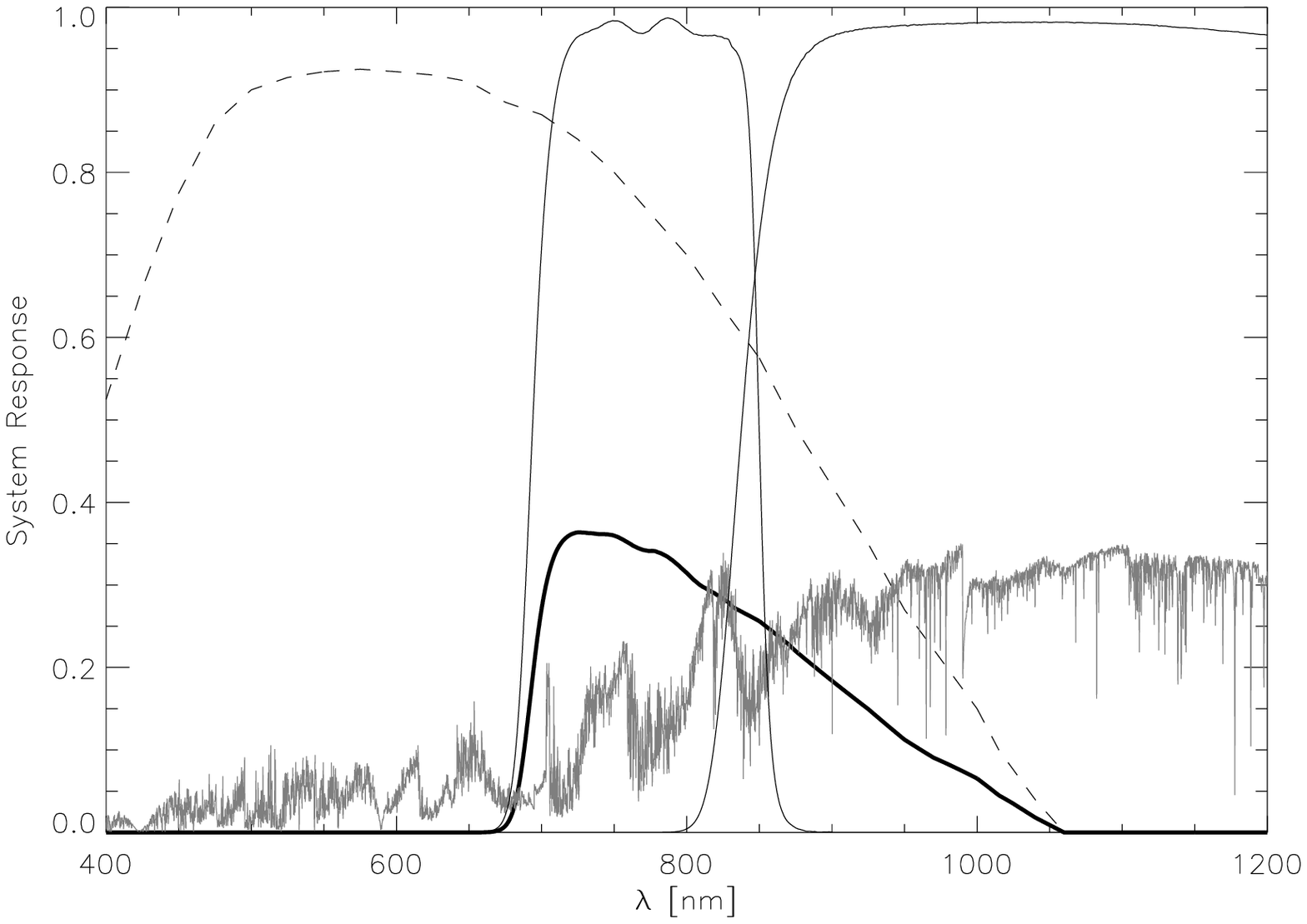}
\caption{The overall system response (thick black) through the $i+z$ filter,
incorporating atmospheric extinction, CCD quantum efficiency, mirror
reflectivity, and an overall 50 \% throughput loss.  
At long wavelengths, the system response is dominated by the CCD quantum efficiency
(dashed).  For
comparison, we show the transmission curves of SDSS $i$ and $z$
filters. In
gray, a NextGen M dwarf spectrum with $T_{\mathrm{eff}} = 3000$ K,
scaled for clarity. }
\label{fig_response}
\end{figure}

With properly scaled synthetic spectra, we can simulate photometry for a
rich variety of transmission functions. We perform our calculations for three
scenarios: through the SDSS $i$ and $z$ filters (as defined by the transmission curves available from the
SDSS DR1 webpage), and through a filter which cuts on and is open beyond $\sim 700$ nm, 
which we'll refer to as the $i+z$ filter.
For optical transmission, we take the square of the reflectivity
curve (i.e. two mirror reflections) measured for a typical aluminum-coated
mirror.  To incorporate atmospheric transmission, we adopt the extinction
coefficients of \citet{Hayes1975}, determined by observation from the Mt.
Hopkins Ridge (altitude 2350 m).  
For CCD response, we assume the quantum efficiency of a typical thinned, back
illuminated CCD camera.  Note that the overall
system response beyond 800 nm is essentially set by the CCD.
From our experience in trying to match simulated photon fluxes to actual
observations of stars with known photometry, we find it prudent to adopt an
additional overall transmission factor of 0.5.  There are a variety of places
where unaccounted for losses of transmission may creep in, e.g. when the QE,
or mirror reflectivity do not meet the manufacturer's specifications.  Note
that we have neglected the reduction in collecting area due to a central
obstruction (e.g. the secondary mirror), but this is more than accommodated for by our assumed loss factor.
The overall system response through the $i+z$ filter is depicted in Fig
\ref{fig_response}, along
with a $T_{\mathrm{eff}} = 3000 K$ synthetic M dwarf spectrum for comparison.

\subsection{Precision and Aperture}

In this section, we look at each of the LSPM M dwarfs, and ask what is the
necessary telescope aperture diameter to achieve a desired precision in a
fixed exposure time.
In this section, we fix the exposure time to 150 seconds, an arbitrarily
chosen exposure time but useful for comparing necessary apertures (later we allow the
exposure time to vary). 
We also set the desired
precision to that which is necessary for a 3 sigma detection, per measurement,
of the transit of a 2-$R_{\earth}$ planet.  
Note that the required precision then varies for each of the
stars, as a function of the estimated stellar radius.   For a 0.33-$R_{\odot}$
M dwarf, a 3 sigma detection requires a precision of 0.001, but for
a 0.10-$R_{\odot}$ M dwarf, it corresponds to a precision of only 0.011.

\begin{figure}[ct1]
\epsscale{1.0}
\plotone{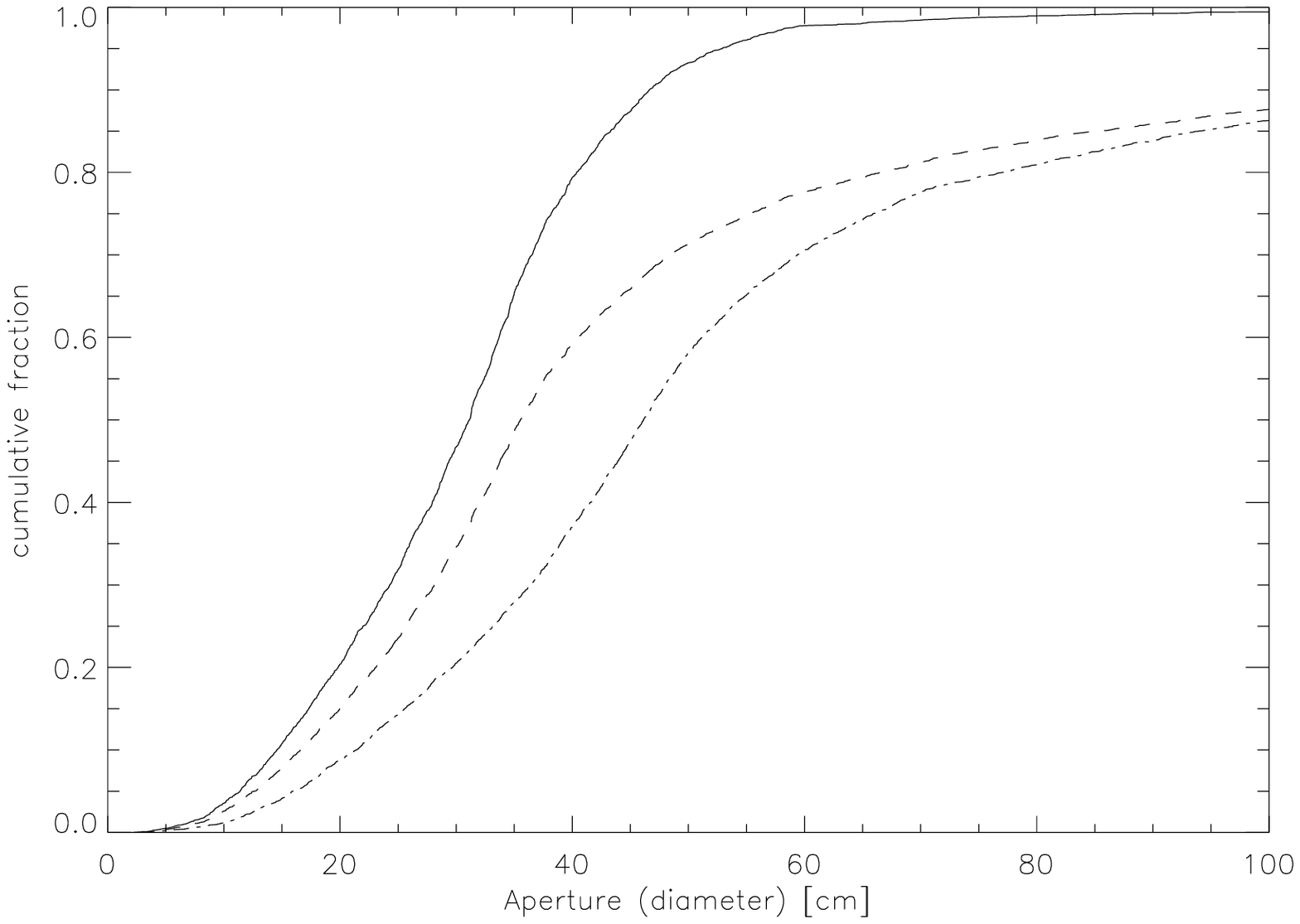}
\caption{Cumulative distribution of LSPM M dwarfs as a function of aperture.
The aperture is that necessary for achieving, in a 150 second integration,
the requisite sensitivity for a 3 sigma detection of a 2-$R_{\earth}$ planet. 
Note that the required sensitivity varies from M dwarf to M dwarf as a
function of stellar radius.  
The dot-dashed curve is for a calculation through the $z$
filter, while the dashed curve is for the $i+z$.  The solid curve is for a
subset of LSPM M dwarfs with estimated radii $ < 0.33 R_{\odot}$
($i+z$ filter).}
\label{cum_ap}
\end{figure}

With this varying precision and fixed exposure time, we have calculated the necessary
aperture for each of the LSPM M dwarfs through the $z$
and $i+z$ filters.  In Figure \ref{cum_ap} we display the cumulative
distribution of LSPM M dwarfs as a function of aperture for the
$z$ and $i+z$ cases.  In comparison to the $z$ filter (dot-dashed curve), it is apparent 
that using the $i+z$ filter (dashed curve) significantly
increases the fraction of stars that meet the desired precision in 150 s.
This is particularly important for apertures in the range 35-40 cm, where use
of $i+z$ more than doubles the number of stars meeting the precision
requirements.

This calculation also drives home a very important point: even though late 
M dwarfs are intrinsically less luminous, and on the mean, fainter than
earlier M dwarfs, this is more than compensated by the relaxed precision
requirements that accompany their smaller radii.  In fact, the stars with the
smallest necessary aperture are dominated by late M dwarfs. In Figure
\ref{cum_ap} we have overplotted (solid curve) the cumulative distribution of stars with
estimated radii $< 0.33 ~R_{\odot}$ (N=1976) as a function of necessary
aperture through the $i+z$ band.
This selection of stars is motivated by the practical difficulty
of achieving a precision better than 0.001 from the ground, which corresponds to the 3 sigma
precision of 2-$R_{\earth}$ planet transiting a 0.33-$R_{\odot}$ star.  
One can see that the aperture requirements for these stars are
quite favorable; 80 \% of the sample can be observed at the requisite
precision in 150 s integrations with telescopes of aperture 40 cm. 

\section{Field of View Requirements} 

Since we target only one star per field, the required size of the
field of view (FOV) is defined
by the need to have an adequate number of calibrating stars. 
For each field, we require the number of photons
received from calibrating stars to be 10 times the number of photons received
from the target M dwarf.  We further require that each calibrating star be
between 0.2 and 1.2 times the brightness of the target star.  Note that these
requirements are relatively strict to give tolerance for the possibility
of variables or other unsuitable stars being among the calibrators.
For each LSPM M dwarf, we determine the size of the smallest square box
(centered on the M dwarf) that meets the calibration requirements.

For this calculation, we query the 2MASS Point Source Catalog
\citep{Cutri2003}, using WCSTools \citep{Mink1998} around the position of each M dwarf.  This
query results in a list of potential calibrating stars.  For each of these
potential calibrators, we transform from 2MASS $J$ and $K_s$ magnitudes 
into estimates of $i$ and $z$ magnitudes.  This transformation is necessary
since the calibrator stars do not, in general, lie in fields covered by the
SDSS survey, and in cases where they do, the SDSS photometry is usually saturated.
The transformation relies on a
polynomial fit to $i-J$ as a function of $J-K_s$ as we now detail.
Our dataset
for the transformation is a sample of cross-matched stars in 2MASS
and the SDSS Photometric Data Release 5 \citep{Adelman-Mccarthy2007} from an
arbitrarily chosen $3^{\circ}$ by $3^{\circ}$ field covered by the SDSS DR5 dataset.  SDSS
photometry was accepted if not flagged as 
SATURATED, EDGE, DEBLENDED\_AS\_MOVING, CHILD, INTERP\_CENTER, or BLENDED.

\begin{figure}[ct1]
\epsscale{1.0}
\plotone{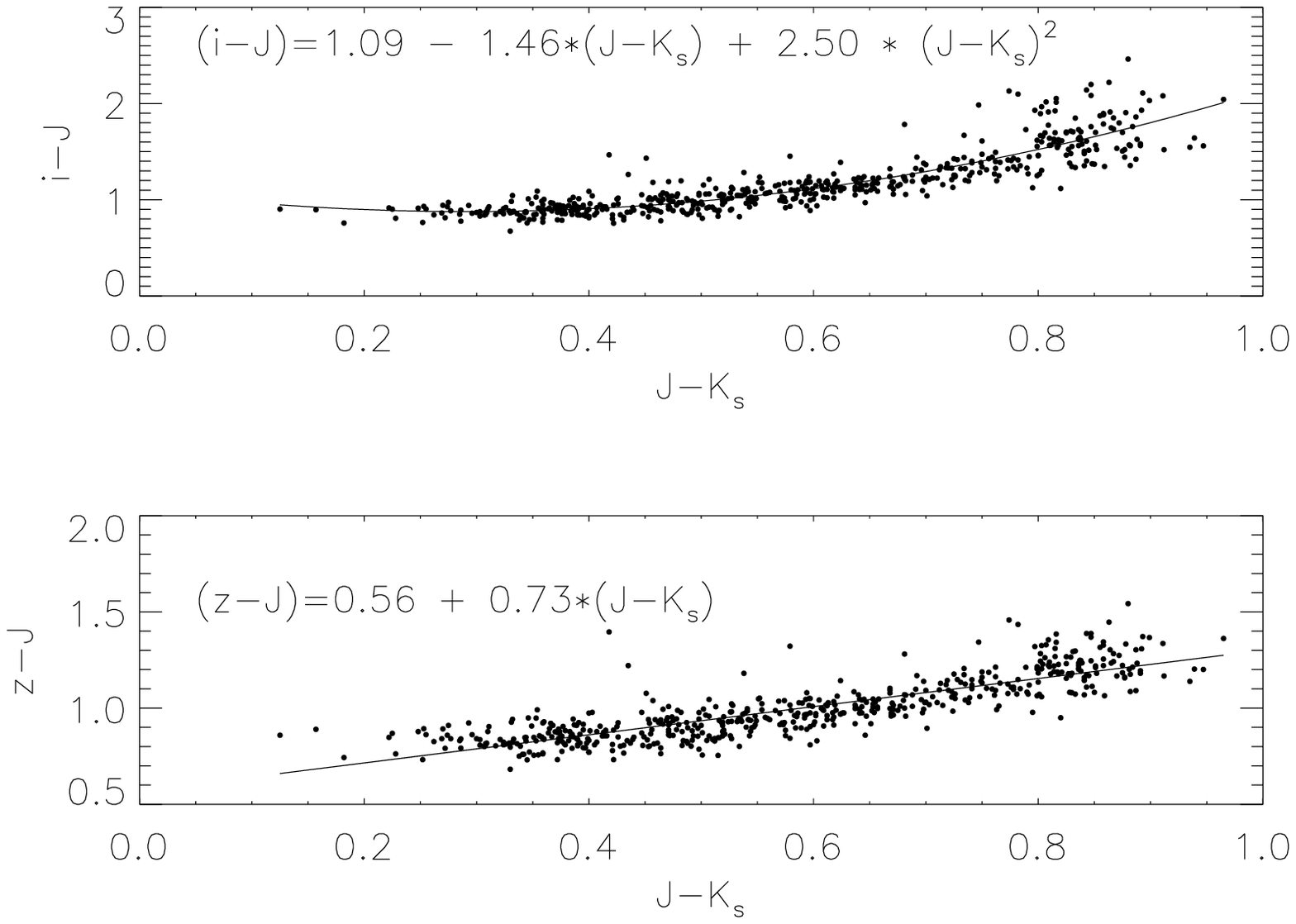}
\caption{Transformation from $J$ and $K_s$ to $i$ (top panel) and $z$
	(bottom panel).  Each point represents a star that has been cross-matched
	in the 2MASS and SDSS surveys from an arbitrarily chosen field covered by
	the SDSS DR5
	dataset.  The displayed best-fit curves and corresponding equations are
	determined by a least squares fit to these data.}
\label{colortrans}
\end{figure}

We derived a quadratic fit to $i-J$ and a linear function fit to $z-J$, each as a
function of $J-K_s$.  The resulting fits, displayed in Figure
\ref{colortrans}, are:
\begin{equation}
i-J =1.09 - 1.46*(J-K_s) + 2.50 * (J-K_s)^2~~~~   0.2 < J-K_s<0.9
\end{equation}
\begin{equation}
z-J =0.56 + 0.73*(J-K_s)	~~~~~~~~~~~~~~~~~~~~~~~~~~~~ 0.2 < J-K_s <0.9
\end{equation}

For the primary M dwarfs, $J-K_s$ is degenerate, so we rely instead on the
$M_J$ vs. $i-J$ and $M_J$ versus $z-J$ relations of \citet{Hawley2002}.
The uncertainty in $M_J$ reaches $\pm 0.6$ mag (dominated by
the uncertainty in distance), but leads to a much smaller uncertainty in $i-J$ ($<
0.3$ mag) and $z-J$ ($<0.2$ mag) because of the small dynamic range of these
colors over the M dwarf sequence.

\begin{figure}[ct1]
\epsscale{1.0}
\plotone{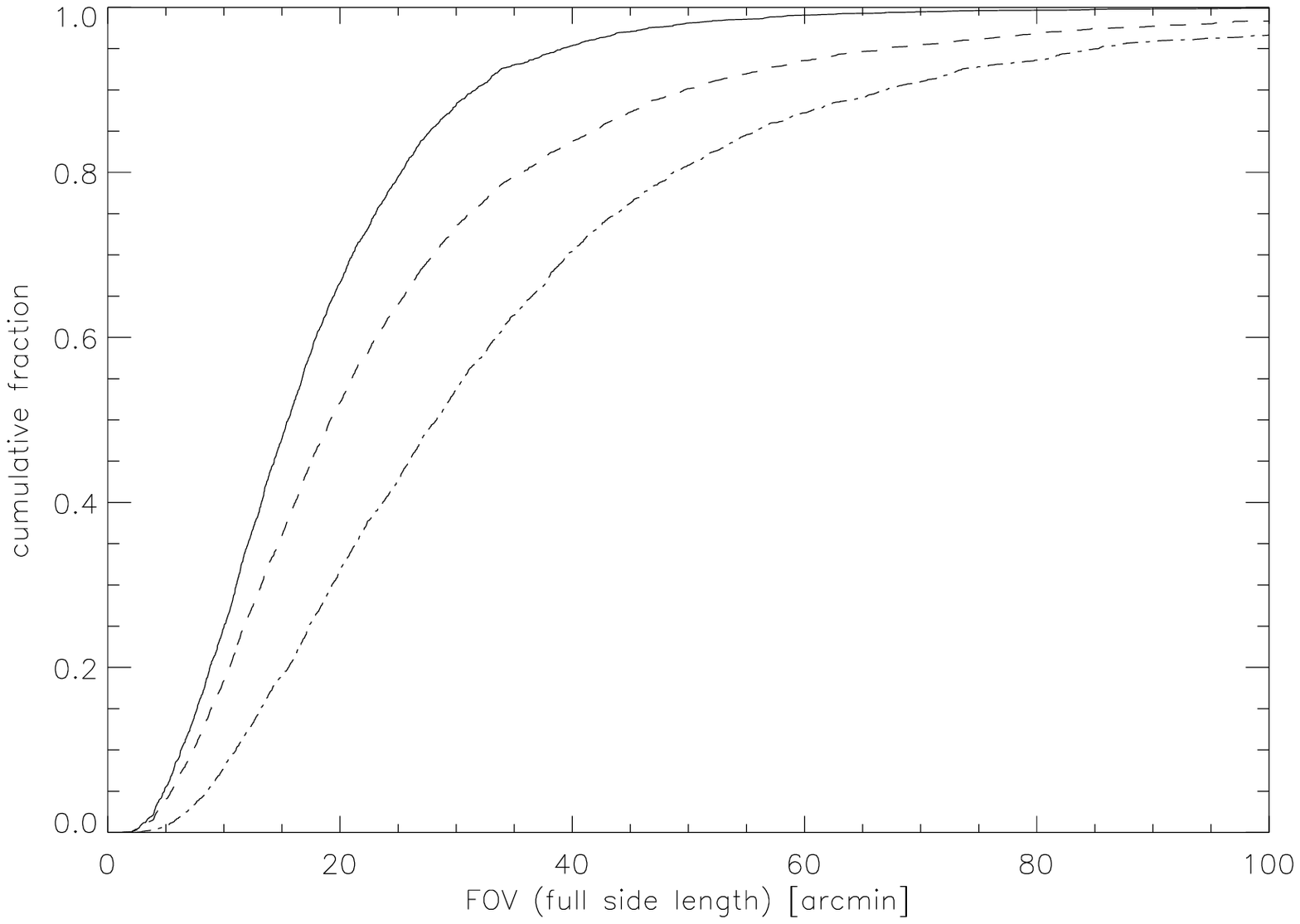}
\caption{Cumulative distribution of LSPM M dwarfs as a function of the necessary field
of view, where we require that the field includes ten times the photon flux
from calibrator stars than from the target M dwarf.  The required FOV is
determined on a star-by-star basis by querying 2MASS for appropriate
calibrating stars around the position of each target M dwarf. The dot-dashed curve is for a calculation through the $z$
filter, while the dashed curve is for the $i+z$.  The solid curve is for a
subset of LSPM M dwarfs with estimated radii $ < 0.33 R_{\odot}$
($i+z$ filter). }
\label{cum_fov}
\end{figure}

We estimate the ratio of photon fluxes in each band by
$10^{-0.4 \Delta i}$ or $10^{-0.4 \Delta z}$, where $\Delta i$, $\Delta z$
are the differences in $i$,$z$ magnitude between target and calibrator.  
To compare photon fluxes
through different bands, one must of course take into account the difference in
relative throughput between each band.  
For these quick estimates, we account for the relative throughput of
$i$ vs. $z$ via the Q factor described in \citet{Fukugita1996}. To
illustrate, a source with equal $i$ and $z$ magnitudes will have approximately
$\mathrm{Q_i}/\mathrm{Q_z}$ ($\approx 2$) more
photons through the $i$ band than through the $z$ band.  With estimates for the
relative number of photons in the calibrators now in hand, we place
successively larger boxes around the target M dwarf, until the calibration
requirement is met.  In Figure \ref{cum_fov}, we show the cumulative 
distribution of all LSPM M dwarfs as a function of the required FOV
for both the $z$ filter (dot-dashed), and $i+z$ filter (dashed).

Some key points arise from this investigation.
An unsurprising one is that brighter targets require significantly larger
fields than fainter targets, due simply to the relative sparseness of brighter
calibrators.  
Another is that using an $i+z$ filter (rather than $z$ alone) saves somewhat  
on the field of view.  
This is again unsurprising since cutting on at a bluer wavelength increases the relative 
number of photons from generally bluer calibrators.  Note that although the
use of the $i+z$ filter adds to the number of M dwarfs observable by our
criteria, it may complicate calibration.  Fringing, for example, may be an issue in this
bandpass with a thinned, back-illuminated CCD.  We expect however, that the
situation will not be too different from the $z$ band observing experiences of
Holman et al. (2007), 
where fringing was apparent, but had little effect
on the photometric precision. In addition, the generally bluer
comparison stars may introduce calibration issues.  We expect that this too
will be surmountable, for example, through the use of color-dependent
extinction corrections.  

The relative faintness of late M dwarfs emerges as a very
favorable characteristic in this calculation.  In Figure \ref{cum_fov}, we have included the
cumulative distribution for the subset of LSPM M dwarfs with estimated radius 
$< 0.33 ~R_{\odot}$ (solid curve) for $i+z$.  This is the same subset of
stars motivated by aperture considerations and described in \S 4.2.

The propitious aspects of these late M dwarfs that arise in both aperture and FOV
considerations are worthy of further investigation.   In Figure 5, we have plotted
the necessary aperture versus necessary field of view for each of the LSPM M
dwarfs, with stars of radius $< 0.33 ~R_{\odot}$ represented by filled circles, and stars of radius
$> 0.33 ~ R_{\odot}$ by open circles.  The criteria for calibrators,
photometric precision, and exposure time are again as described above.  
We see that the late M dwarfs occupy a locus in aperture-FOV
space that is very fortunate from an instrumental design standpoint.

\begin{figure}[ct1]
\epsscale{1.0}
\plotone{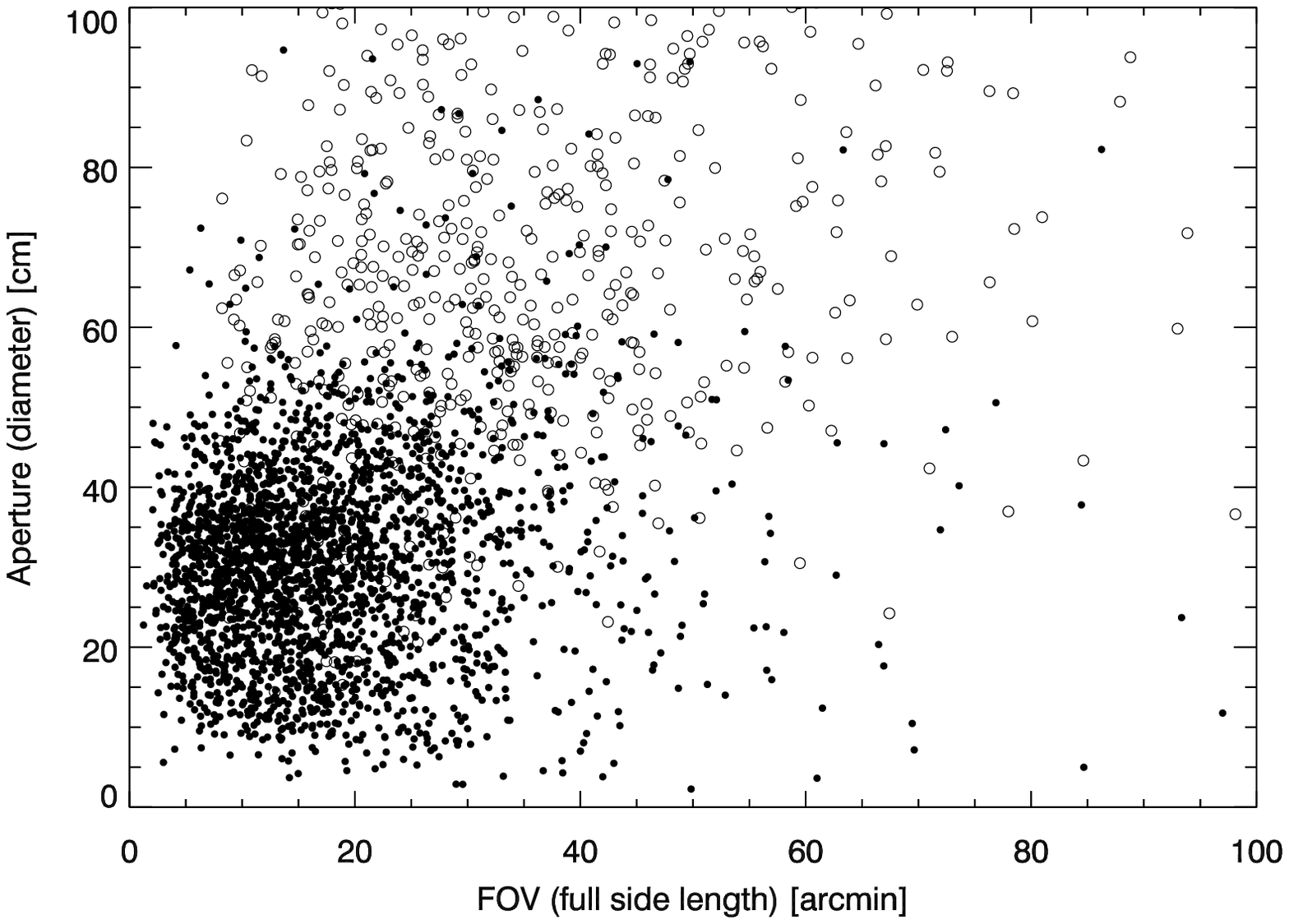}
\caption{ Necessary aperture vs. necessary FOV for LSPM M dwarfs.
Stars with radius $ > 0.33 ~ R_{\odot} $ are represented by open circles,
while stars with radius $ < 0.33 ~ R_{\odot}$ are represented by filled
	circles.  The required precision (for aperture calculation) is that
	necessary to achieve a 3 sigma detection of a transiting 2 $R_{\earth}$
	planet in a 150 second integration through the $i+z$ filter.  The range of
	the x and y axes match that of the radius $ < 0.33 ~ R_{\odot}$ stars,
	while 30 \% of the radius $ > 0.33 ~ R_{\odot} $ M dwarfs require more
		than a 100 cm aperture, and thus fall above the plot limits.
	}
\label{ap_fov}
\end{figure}

\section{Survey Duration and Number of Telescopes}

In this section we estimate the amount of telescope time necessary for a
successful MEarth Project.  Our conclusion is framed in units of
\emph{telescope-years}, reflecting the fact that the survey duration and the number of
telescopes is a trade-off.
Our calculation is tailored to the characteristics of 
2-$R_{\earth}$-sized planets orbiting in the habitable zones of the host
stars. 

The outline of the calculation is as follows. We 
first determine the fraction of a telescope's time, f$_{\mathrm{tel}}$ that must be devoted
to each star to guarantee the temporal coverage necessary for catching
transits of habitable zone planets.  We then calculate the number of
observing nights, N$_{\mathrm{nights}}$, that are necessary until one can be 90
\% confident that at least one transit would have fallen during an observation
session.  
The effective number of telescope nights,
N$_{\mathrm{eff}}$ is then N$_{\mathrm{eff}} = $ N$_{\mathrm{nights}}
\times$ f$_{\mathrm{tel}}$.  The total effective number of telescope nights is
determined by calculating and summing N$_{\mathrm{eff}}$ over the list of
stars to survey.

\begin{figure}[ct1]
\epsscale{1.0}
\plotone{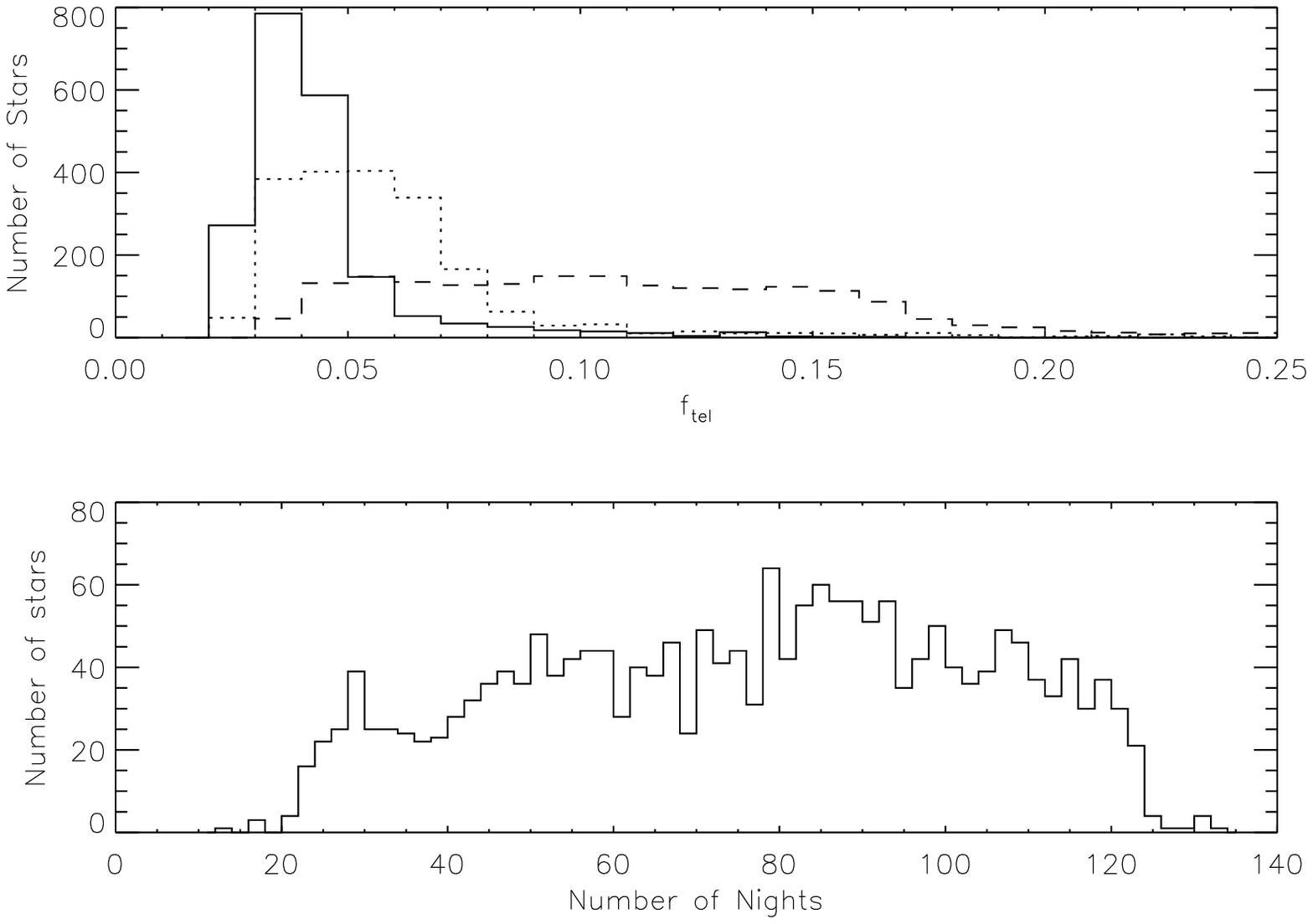}
\caption{\emph{Top:} Histogram of the late M dwarf sample, binned by f$_{\mathrm{tel}}$
, the fraction of telescope's time devoted to the star.  The dashed curve
give results for 20 cm aperture, dotted for 35 cm, and solid for 50 cm.
\emph{Bottom:}  Histogram of the late M dwarf sample, binned by N$_{\mathrm{nights}}$, the number of
observing nights required to be 90 \% confident that at least one transit
event from a habitable zone-orbiting planet would have fallen during an
observational session.
}
\label{tel_histo}
\end{figure}

The calculation of f$_{\mathrm{tel}}$ is done as follows:
1) For each star, and for a given
aperture, we determine the necessary exposure time to achieve a 3 sigma detection
of a 2 $R_{\earth}$ planet. In addition, we assume an overhead time of 60 s to account
for time spent slewing between targets.  
2) We determine the transit duration using the
inferred stellar parameters from section 3.  We assume a circular orbit
at a distance from the star for which the planet receives the same stellar
flux as the Earth.  We further assume a
mid-latitude transit, which leads to a transit duration of 0.866 times that of
an equatorial transit.
3) We require at least 2 visits to the star per transit duration. This then sets
the cadence to 2/transit duration (cycles per unit time). 
We impose a minimum cadence of one
visit per 60 minutes, to ensure that we have sufficient temporal coverage to
catch shorter transit duration planets (i.e. planets interior of the
habitable zone).  
4) Finally, the f$_{\mathrm{tel}}$ dedicated to a given star is given by f$_{\mathrm{tel}}$ = cadence $\times$ (exposure time + overhead). 
In the top panel of Figure \ref{tel_histo}, we display a histogram of the
sample of 1,976 late M
dwarfs, binned by f$_{\mathrm{tel}}$.  These histograms give a sense of what
fraction of a telescope's time must be devoted to individual M dwarfs, when those stars are
being actively observed.  
f$_{\mathrm{tel}}$ is calculated as described above for aperture diameters of
20 cm (dashed), 35 cm (dotted), and 50 cm (solid).  One can see that for 20 cm apertures, a
significant fraction of stars would require more than 10 \% of the telescope's
time while actively being observed.  For 35 and 50 cm apertures, a typical star
requires $\sim$ 5\% and $\sim$ 3\%, respectively, of a telescope's time.

Our calculation for N$_{\mathrm{nights}}$ involves simulations similar to
those performed in `window function' calculations that are common in the literature 
(see, for example, Pepper et al 2005). \nocite{Pepper2005}  The major
difference is that here we require only one transit.  This is justified because we
anticipate reducing photometry in real-time, so that transits events can identified
and confirmed while still in progress.  We inject transit signals over an extensive grid of possible phases, with periods 
assigned to each star corresponding to planets in habitable zone orbits.  
For the purposes of simulation, we assume observational seasons of 60 nights with 9 hours of observing each night.  
We randomly knock out 50 \% of observing nights to account for weather effects.
In practice, most stars in the sample are visible from Mt. Hopkins (latitude
31.6$^{\circ}$ N)  for more than 60 nights each year (neglecting weather) and are visible for less than 9 hours each night, but
9 hours $\times$ 60 roughly represents the number of hours
a typical star is visible over the course of a season. 
To avoid skewed results from periods near integer or half-integer number of days (which are known to show
resonances in detection probability),
we simulate over a uniform range of possible periods for each star.  The upper end of this range of periods
is defined by planets receiving the same stellar flux as Earth, and at the lower end by planets with 
an equilibrium temperature of 290 K (assuming a wavelength-integrated Bond
albedo of 0.3).  We determine N$_{\mathrm{nights}}$ by
requiring that 90 \% of transit signals are recovered.
In the bottom panel of Figure \ref{tel_histo}, we give a histogram for the
total number of nights during which
each star must observed. 

\begin{figure}[ct1]
\epsscale{1.0}
\plotone{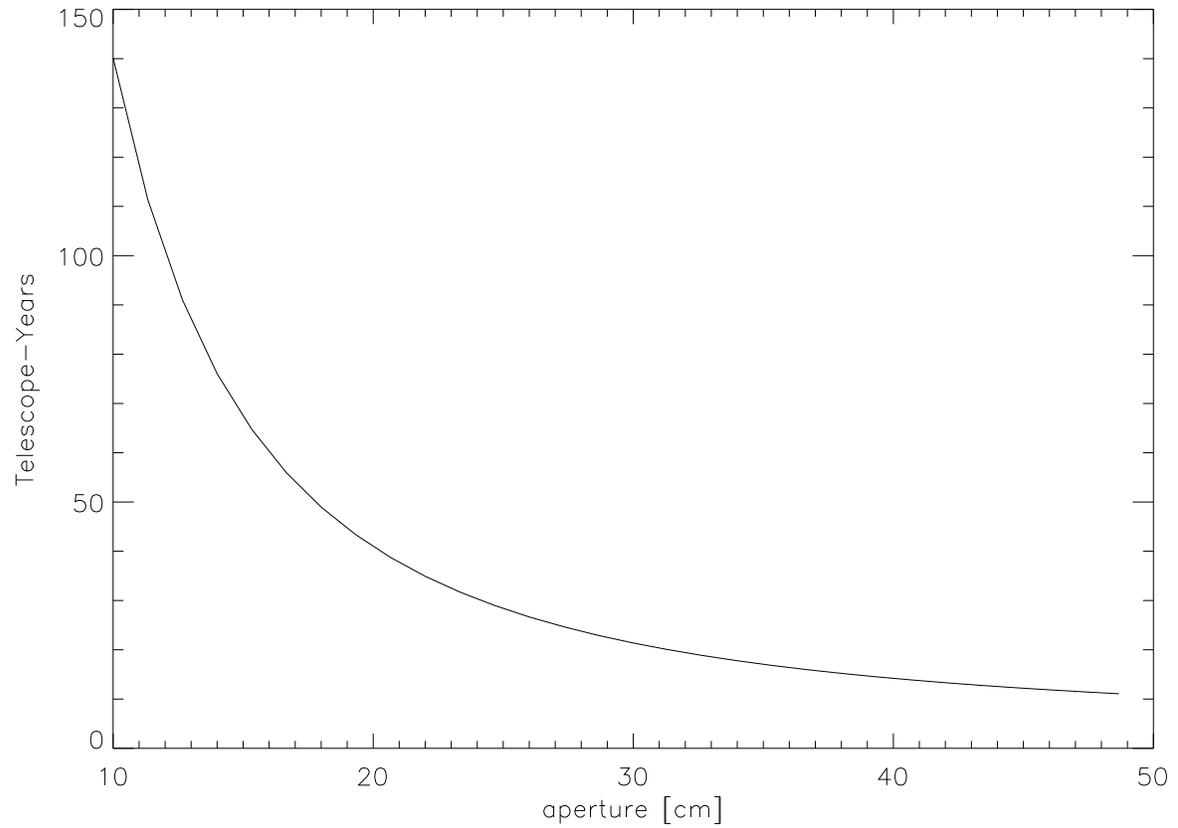}
\caption{Number of telescope-years required to survey the sample of 1,976
late M dwarfs, as a function of aperture.  This is calculated by summing
f$_{\mathrm{tel}} \times $N$_{\mathrm{nights}}$ over each M dwarf, where f$_{\mathrm{tel}}$ is the fraction of telescope's time devoted
to the star, and N$_{\mathrm{nights}}$ is the number of
observing nights required to be 90 \% confident that at least one transit
event from a habitable zone-orbiting planet would have fallen during an
observational session.
}
\label{eff_nights}
\end{figure}

The only remaining issue is to select which M dwarfs to sum N$_{\mathrm{eff}}$
over.  The stars with the optimal N$_{\mathrm{eff}}$ are the
coolest M dwarfs, for which the periods of habitable zone orbits are shortest.  
The previously described sample of 1,976 late M dwarfs with radius $< ~ 0.33
R_{\odot}$ are once again very appropriate under this consideration.  In
Figure \ref{eff_nights}, we have summed N$_{\mathrm{eff}}$ over these M dwarfs as a function of
aperture.  The effect of increasing the aperture size is to decrease the
integration times required to each star, and hence to decrease the fraction of
its time that a
telescope must devote to each star.  At $\sim$ 30 cm, the marginal benefit
of adding more aperture diminishes, simply because at this aperture,
overhead time spent slewing between targets begins to dominate over the
actual time spent integrating on targets. 

At 30 cm, N$_{\mathrm{eff}}$=22.1 telescope-years. Thus ten such telescopes could survey the sample of 1,976
late M dwarfs in 2.2 years.  Note that simply adding more
telescopes to the network does not necessarily reduce the survey completion
time.  For example a significant fraction of the stars in the late M dwarf
sample require a time baseline of more than 90 nights to achieve 90 \%
confidence that a transit event would have occurred during an observational session
(see Figure \ref{tel_histo}).
If many of these stars are only visible 30 good weather nights a year, then
observations of these stars must be spread out over 3 years, regardless of the
amount of telescope time one devotes to each star.

\section{Conclusions and Discussion}

We have investigated the design requirements for the MEarth Project, a
survey conceived to monitor Northern Hemisphere M dwarfs for transits of habitable
planets, with a sensitivity to detect planets down to a radius of $2 R_{\earth}$.
In our investigation, 1,976 late M dwarfs ($R < 0.33 R_{\odot}$) emerged as the most favorable survey
targets, initially for reasons related to photometric precision. 
Despite their relative faintness, it is easier to achieve the required
sensitivity to detect a given planet because their small radii lead to deep transit signals.  In
consideration of the required field-of-view, late M dwarfs once again arose as
the most favorable survey targets-- in this case \emph{because} of their relative
faintness.  A final investigation into the amount of telescope time required to
achieve transit detections of habitable planets again favored late M dwarfs,
because of the short periods of habitable zone orbits. 

Because of
the increased geometric probability of transit for habitable planets around
the late M dwarfs, the constraints on the occurrence rate of such planets
are correspondingly tighter.  We can perform an ex post facto analysis on the sample
of 1,976 late M dwarfs with an estimated $R < 0.33 R_{\odot}$, using their
individual inferred stellar radii and the estimated semi-major axes of planets in
their habitable zones, and assuming a recovery rate of 90 \% for planets
transiting from the habitable zone.  For this sample of stars, a lack of any transit detections of
habitable planets would lead to an upper limit (99 \% confidence) of 17 \% for
the occurrence of such
planets.  For a true occurrence rate of 10 \% for habitable planets
(larger than 2 $R_{\earth}$), the expected yield would be 2.6 such planets.
We note that for even closer planets, such as the hot Neptune transiting
GJ 436, the expected yield is significantly larger, and thus, in their own
right could justify a project to monitor this many M dwarfs.  We also note
that our sample of M dwarfs include 450 stars with an estimated $R < 0.17
R_{\odot}$, to which the MEarth network could be sensitive to transiting
planets as small as 1 $\rearth$.  To achieve this sensitivity, the exposure times
for these 450 stars would need to be increased by roughly a factor of 4 compared to the exposure
times calculated in this paper.

Once built, the MEarth network of robotic telescopes will be able to survey
the 1,976 late M dwarfs in 22 telescope-years, if equipped with 30 cm aperture
telescopes, using the 'i+z' filter described in section 4. 
From an instrumental viewpoint, successfully observing this sample
of M dwarfs is challenged by the $\sim$ 10 \% of
this sample (see Figure \ref{cum_fov}) for which the estimated field of view
requirements are greater than 30 arcmin by 30 arcmin.  Possible
solutions to this challenge worth exploring include, for example, the addition of a wide angle-node
to the MEarth network, or simply using a large-format camera.  It may also be possible to accommodate these
stars by using custom field orientations in order to grab extra
calibrating stars, or to relax the conservative calibrating criteria that we
assumed in our field of view calculations.      

This study has confirmed the status of nearby
late M dwarfs as bearers of the lowest hanging fruit in the search for
habitable rocky planets.  Excitingly, these
stars remain largely unexplored: Since late M dwarfs are very faint at the
visible wavelengths at which iodine provides reference lines, they are
not accessible to current radial velocity planet searches. 
Besides the search for transiting planets, a plan to photometrically monitor this many M
dwarfs represents a large step forward in the study of the intrinsic
variability, and long-term activity of M dwarfs.  The identification and monitoring of spotted stars, for example, will
be useful to future, near IR radial velocity observational programs which will
be compromised by the radial velocity jitter and spurious signals that might
result.
\linebreak

We gratefully acknowledge funding for this project from the David and Lucile
Packard Fellowship for Science and Engineering.
We would like to thank Andrew Szentgyorgyi, David Latham, Matt Holman 
and Cullen Blake for helpful comments, and an anonymous referee for thoughtful
comments and helpful recommendations.
This publication makes use of data products from the Two Micron All Sky
Survey, which is a joint project of the University of Massachusetts and the
Infrared Processing and Analysis Center/California Institute of Technology,
funded by the National Aeronautics and Space Administration and the National
Science Foundation.

\bibliographystyle{apj}

\begin{deluxetable}{lcccc} 
\setlength{\tabcolsep}{0.1in}
\tablecaption{LSPM M dwarf Parameters\tablenotemark{a} }
\tablecolumns{7}
\tablewidth{0pt}
\tabletypesize{\footnotesize}
\tablehead{
  \colhead{Sp. Type \tablenotemark{b}}							&
  \colhead{$N$}									&
\colhead{$R$ [$R_{\sun}$]} 						&
  \colhead{$M$ [$M_{\sun}$]}					&
  \colhead{$J$}					
}
\startdata
M0	&	129		&	.69	&	.68	&	7.58			    \\  %
M1			&	128		&	.62	&	.61	&   7.80                \\  %
M2			&	225		&	.55	&	.54	&   8.10                \\  %
M3			&	419		&	.44	&	.43	&   8.64                \\  %
M4			&	890		&	.33	&	.32	&   9.44                \\  %
M5			&	1043	&	.24	&	.22	&   10.21               \\  %
M6			&	348		&	.15	&	.13	&   11.15               \\  %
M7 \& M8	&	114		&	.12	&	.09	&   12.06        		 \\  %
                                                         
\enddata
\tablenotetext{a}{Mean Values for different radius bins}
\tablenotetext{b}{Spectral type estimated by a fit to Leggett (2000) data as
function of $V-K$ color.}
\end{deluxetable}

\end{document}